\documentstyle[prl,aps]{revtex}
\begin{document}

\input{epsf.tex}
\epsfverbosetrue
\title{Spatial Optical Solitons due to Multistep Cascading}

\author{Yuri S. Kivshar and Tristram J. Alexander}

\address{Optical Sciences Center, Research School of Physical
Sciences and Engineering \\ Australian National University,
Canberra ACT 0200, Australia}

\author{Solomon Saltiel}

\address{Faculty of Physics, University of Sofia, Sofia 1164, Bulgaria}

\maketitle

\begin{abstract}
We introduce a novel class of parametric optical solitons supported
simultaneously by two second-order nonlinear cascading processes, {\em
second-harmonic generation} and {\em sum-frequency mixing}.  We obtain,
analytically and numerically, the solutions for three-wave spatial solitons
and show that the presence of an additional cascading mechanism can change
dramatically the properties and stability of two-wave quadratic solitary
waves.
\end{abstract}

\vspace{10mm}

As is known, optical cascaded nonlinearities due to parametric wave mixing
can lead to a large nonlinear phase shift and spatial solitary  waves,
resembling those for a Kerr medium \cite{steg}.   However, solitary waves
supported by cascaded nonlinearities demonstrate much richer dynamics due
to nonintegrability  of governing nonlinear equations and, unlike solitons
of the Kerr nonlinearity, the quadratic solitons can become unstable in a
certain narrow region of their parameters \cite{pelin}.

In this Letter we introduce {\em a novel class of parametric spatial
solitons} supported simultaneously by two nonlinear quadratic (or
$\chi^{(2)}$) optical  processes: second-harmonic generation (SHG) and
sum-frequency mixing (SFM).  As has been recently shown  by Koynov and
Saltiel \cite{koy_sal} for continuous waves, under the condition that the
two wave-mixing processes are nearly phase matched, the presence of
multistep cascading  leads to a four fold reduction of the input intensity
required to achieve a large nonlinear  phase shift.  Here, we demonstrate
that the multistep cascading can lead  to {\em a new type of parametric
solitons}. Introducing a third wave generated via a SFM process, we find
that it can alter both the general properties and stability of the two-wave
$\chi^{(2)}$ spatial solitons. Moreover, we reveal the existence of a new
type of the so-called {\em quasi-soliton}, that appear for a negative
mismatch of the SFM process.

To introduce the model of multistep cascading,  we consider the fundamental
beam with frequency $\omega$ entering a noncentrosymmetric nonlinear medium
with a $\chi^{(2)}$  response.  As a first step, the second-harmonic wave
with frequency $2\omega$  is generated via the SHG process.  As a second
step,  we expect the generation  of higher order harmonics due to SFM, for
example, a third harmonic ($\omega +  2\omega = 3\omega$) or even fourth
harmonic ($2\omega + 2\omega = 4\omega$) \cite{akhmanov}.  When both such
processes are nearly phase matched,  they can lead, via down-convertion, to
a large nonlinear phase shift of  the fundamental wave \cite{koy_sal}.
Additionally, as we demonstrate in this  paper, the multistep cascading can
support {\em a novel type of three-wave spatial solitary waves} in a
diffractive $\chi^{(2)}$ nonlinear medium, {\em multistep cascading
solitons}.

We start our analysis with the reduced amplitude equations derived in the
slowly varying  envelope approximation with the assumption of zero
absorption of all  interacting waves (see, e.g., Ref. \cite{koy_sal}).
Introducing the effect of  diffraction in a slab waveguide geometry, we
obtain
\begin{equation}
\label{physeqns}
\begin{array}{l}
{\displaystyle 2ik_{1}\frac{\partial A_{1}}{\partial z} +
\frac{\partial^{2} A_{1}}
{\partial x^{2}} + \chi_{1}A_{3}A_{2}^{\ast}e^{-i\Delta k_{3}z} + \chi_{2}A_{2}
A_{1}^{\ast}e^{-i\Delta k_{2}z} = 0, } \\*[9pt]
{\displaystyle 4ik_{1}\frac{\partial A_{2}}{\partial z} +
\frac{\partial^{2} A_{2}}
{\partial x^{2}} + \chi_{4}A_{3}A_{1}^{\ast}e^{-i\Delta k_{3}z} + \chi_{5}
A_{1}^{2}e^{i\Delta k_{2}z} = 0, } \\*[9pt]
{\displaystyle 6ik_{1}\frac{\partial A_{3}}{\partial z} +
\frac{\partial^{2} A_{3}}
{\partial x^{2}} + \chi_{3}A_{2}A_{1}e^{i\Delta k_{3}z} = 0, } \\*[9pt]
\end{array}
\end{equation}
where $\chi_{1,2} = 2k_1 \sigma_{1,2}$, $\chi_3 = 6k_1 \sigma_3$, and
$\chi_{4,5} = 4k_1 \sigma_{4,5}$, and the nonlinear coupling coefficients
$\sigma_k$ are proportional to the elements of the second-order
susceptibility tensor which we assume to satisfy the following relations
(no dispersion), $\sigma_3 = 3\sigma_1$, $\sigma_2 = \sigma_5$, and
$\sigma_4 = 2\sigma_1$.

In Eqs. (\ref{physeqns}),  $A_{1}$,$A_{2}$ and $A_{3}$ are the complex
electric field envelopes of the  fundamental harmonic (FH), second harmonic
(SH), and third harmonic (TH),  respectively, $\Delta k_{2} = 2 k_1 - k_2$
is the wavevector mismatch for the  SHG process,  and $\Delta k_{3} = k_1+
k_2 - k_3$ is the wavevector mismatch  for the SFM process.  The subscripts
`1' denote the FH wave, the subscripts `2'  denote the SH wave, and the
subscripts `3', the TH wave.  Following the technique earlier employed in
Refs.  \cite{ol_bur}, we look for stationary solutions of Eq.
(\ref{physeqns}) and introduce the normalised  envelope $w(z,x)$, $v(z,x)$,
and $u(z,x)$ according to the relations,
\begin{equation}
\label{substitute}
A_{1} = \frac{\sqrt{2}\beta k_{1}}{\sqrt{\chi_{2}\chi_{5}}}e^{i\beta z} w,
\;\;  A_{2} = \frac{2 \beta k_{1}}{\chi_{2}}e^{2i\beta z + i\Delta k_{2} z}
v, \;\; A_{3} = \frac{\sqrt{2\chi_{2}}\beta
k_{1}}{\chi_{1}\sqrt{\chi_{5}}}e^{3i\beta z + i\Delta k z} u,
\end{equation}
where $\Delta k \equiv \Delta k_{2} + \Delta k_{3}$.  Renormalising the
variables as  $z \rightarrow z/\beta$ and $x \rightarrow
x/\sqrt{2\beta k_{1}}$, we finally obtain a system of coupled equations,
\begin{equation}
\label{normal}
\begin{array}{l}
{\displaystyle i\frac{\partial w}{\partial z} + \frac{\partial^{2}
w}{\partial x^{2}}  - w + w^{\ast}v + v^{\ast}u = 0,} \\*[9pt]
{\displaystyle 2i \frac{\partial v}{\partial z} + \frac{\partial^{2}
v}{\partial x^{2}} - \alpha v + \frac{1}{2} w^{2} + w^{\ast}u = 0,} \\*[9pt]
{\displaystyle 3i\frac{\partial u}{\partial z} + \frac{\partial^{2}
u}{\partial x^{2}} - \alpha_{1}u + \chi vw = 0,} \\*[9pt]
\end{array}
\end{equation}
where $\alpha = 2(2\beta + \Delta k_{2})/\beta$ and $\alpha_{1} = 3(3\beta
+ \Delta k)/ \beta$ are two dimensionless parameters that characterise the
nonlinear phase matching  between the parametrically interacting waves.
Dimensionless material parameter $\chi \equiv  \chi_{1}\chi_{3}/ \chi_{2}^2
=
9 (\sigma_1/\sigma_2)^2$ depends on the type of phase matching, and it can
take different values of order of one.  For example, when both SHG and SFM
are due to quasi-phase matching (QPM), we have $\sigma_j = (2/\pi
m)(\pi/\lambda_1 n_1)  \chi^{(2)} [\omega; (4-j)\omega; - (3-j)\omega]$,
where $j =1,2$. Then, for the first-order $(m =1)$  QPM processes (see,
e.g., Ref. \cite{pfister}), we have $\sigma_1 = \sigma_2$,  and therefore
$\chi=9$. When SFM is due to the third-order QPM process (see, e.g., Ref.
\cite{baldi}), we should take $\sigma_1 = \sigma_2/3$, and therefore
$\chi=1$. At last, when SFM is the fifth-order QPM process, we have
$\sigma_1 = \sigma_2/5$ and $\chi=9/25$.

Dimensionless equations (\ref{normal}) present a fundamental model for
three-wave  multistep cascading solitons in the absence of walk-off.
Additionally to the type I SHG solitons (see, e.g., Refs \cite{ol_bur}),
the multistep cascading solitons involve the phase-matched SFM interaction
($\omega + 2\omega = 3\omega$) that generates a third harmonic  wave.  If
this latter process is not phase-matched, we should consider $\alpha_{1}$
as a large parameter, and then look for solutions of Eq. (\ref{normal}) in
the form of an asymptotic series
in $\alpha_{1}$.  Substituting $w = w_{0} + \varepsilon w_{1} + \ldots$, $v
= v_{0}  + \varepsilon v_{1} + \ldots$ and $u = \varepsilon u_{1}$, where
$\varepsilon \equiv  \alpha_{1}^{-1}$,  we find $u_{1} \approx \chi vw$,
and the system (\ref{normal}) reduces to a model of competing nonlinearities,
\begin{equation}
\label{reduced}
\begin{array}{l}
{\displaystyle i\frac{\partial w}{\partial z} + \frac{\partial^{2} w}
{\partial x^{2}}- w + w^{\ast}v + \varepsilon\chi |v|^{2}w = 0,} \\*[9pt]
{\displaystyle 2i\frac{\partial v}{\partial z} + \frac{\partial^{2}
v}{\partial x^{2}} - \alpha v + \frac{w^{2}}{2} + \varepsilon \chi |w|^{2}v
= 0.} \\*[9pt]
\end{array}
\end{equation}
In the limit $\varepsilon \rightarrow 0$, Eqs. (\ref{reduced}) coincide
with the model  of two-wave solitons due to the type I SHG earlier analysed
in Refs. \cite{ol_bur}.

For smaller $\alpha_1$, the system (\ref{normal}) cannot be reduced to Eq.
(\ref{reduced}),  and its two-parameter family of localised  solutions
consists of three mutually coupled waves. It is interesting to note that,
similar to the case of nondegenerate three-wave mixing  \cite{malomed},
Eqs. (\ref{normal}) possess an exact solution.  To find it, we make
a substitution $w = w_{0}\, {\rm sech}^{2}(\eta x)$, $v = v_{0} \,{\rm
sech}^{2}(\eta x)$ and  $u = u_{0}\, {\rm sech}^{2}(\eta x)$,  and obtain
 unknown parameters from the following algebraic  equations
\begin{equation}
\label{exactsol}
w_{0}^{2} = \frac{9v_{0}}{3+4\chi v_{0}}, \;\;\; 4\chi v_{0}^{2} + 6v_{0}
=9 , \;\;\; u_{0} = \frac{2}{3} \chi w_{0}v_{0},
\end{equation}
valid for $\eta = \frac{1}{2}$ and $\alpha = \alpha_{1} = 1$. Equations
(\ref{exactsol}) have two solutions corresponding to {\em positive} and
{\em negative}  values of the amplitude.  This indicates a possibility of
multivalued solutions, even within the class of exact solutions.

In a general case, three-wave solitons of Eqs. (\ref{normal}) can be found
only numerically.  Figures 1(a) and 1(b)
present two examples of solitary waves for different sets of the mismatch
parameters  $\alpha$ and $\alpha_{1}$.  When $\alpha_{1} \gg 1$ [see Fig.
1(a)], which corresponds to an unmatched SFM process, the amplitude of the
third harmonic is small, and it vanishes for  $\alpha_{1} \rightarrow
\infty$ according to the asymptotic solution of Eq. (\ref{reduced})
discussed above.

To summarise different types of three-wave solitary waves, in  Fig. 2 we
plot the dependence of the total soliton power defined as
\begin{equation}
\label{power}
P = \int^{+\infty}_{-\infty} dx \left( |w|^2 + 4 |v|^2 + \frac{9}{\chi}
|u|^2 \right),
\end{equation}
on the mismatch parameter $\alpha_1$, for fixed $\alpha=1$. It is clearly
seen that for some
values of $\alpha_1$ (including the exact solution at $\alpha_1 =1$ shown
by two filled circles), there exist {\em two different branches} of
three-wave solitary waves, and only one of those branches approaches, for
large values of $\alpha_1$,  a family of two-wave solitons of the cascading
limit
(Fig. 2, dashed).  The slope of the branches changes from negative (for
small $\alpha_1$) to positive (for large $\alpha_1$), indicating a possible
change of the soliton stability. However, the soliton stability should be
defined in terms of physical parameters, and in the case of two-parameter
solitons as we have here, the stability threshold is determined by a
certain integral determinant condition,  similar to that first derived for
the three-wave mixing problem \cite{trillo}.

Ratios of the maximum amplitudes of the soliton components for the
three-wave solitons of the lower branch in the model (\ref{normal}) are
presented in  Fig. 3, where the upper dashed curve is the asymptotic limit
of two-wave solitons for $\alpha_1 \rightarrow \infty$.  Soliton solutions
of the second (upper) branch in Fig. 2  correspond to large values of the
total power and they have been verified numerically to be unstable.

The analysis of the asymptotics for Eqs. (\ref{normal}) suggests that
localised solutions should not occur for $\alpha_1 <0$. However, we reveal
the existence of an extended class of very robust localised solutions which
we classify as  `quasi-solitons' \cite{zakh}, solitary waves with
small-amplitude oscillating tails. In principle, such solitons are known in
one-component models (see, e.g., Ref. \cite{KdV}) but here the nonvanishing
tails appear only due to  a resonance with the third-harmonic field [see
Fig. 4(a)]. Such solitons are expected to be weakly unstable, and this is
indeed demonstrated in Fig. 4(b) for rather long propagation distances.

Existence of quasi-solitons for any value of negative phase-matching with a
higher-order harmonic field indicates that all two-wave quadratic solitons
{\em can become unstable} due to an additional SFM process. This is
confirmed in Figs. 4(c,d) where we present the results of numerical
simulations of the dynamics of an initially launched two-wave soliton for
two cases, positive and negative phase-matching of a SFM process. For
$\alpha_1 > 0$
[see Fig. 4(c)], a very small harmonic ($v_{\rm max} \approx 0.1$) is
generated and the initial two-component beam converges to a three-wave
soliton. In contrast, for $\alpha_1 <0$ [see Fig. 4(d)], the input beam
decays rapidly into radiation and diffracting harmonic fields.

In conclusion, we have investigated, analytically and numerically,
multistep cascading and nonlinear beam propagation in a diffractive optical
medium and introduced  a novel type of three-wave parametric  spatial
optical solitons, multistep cascading solitons.  The detailed analysis of
the soliton stability, the effect of walk-off, higher-dimensional and
spatio-temporal effects are  possible directions of the future research.

The authors are indebted to K. Koynov, R. Schiek, and E. Kuznetsov  for
useful discussions. The work has been partially supported by the Australian
Photonics Cooperative Research Centre and the Australian Research Council.

\begin{figure}
\setlength{\epsfxsize}{15.0cm}
\mbox{\epsffile{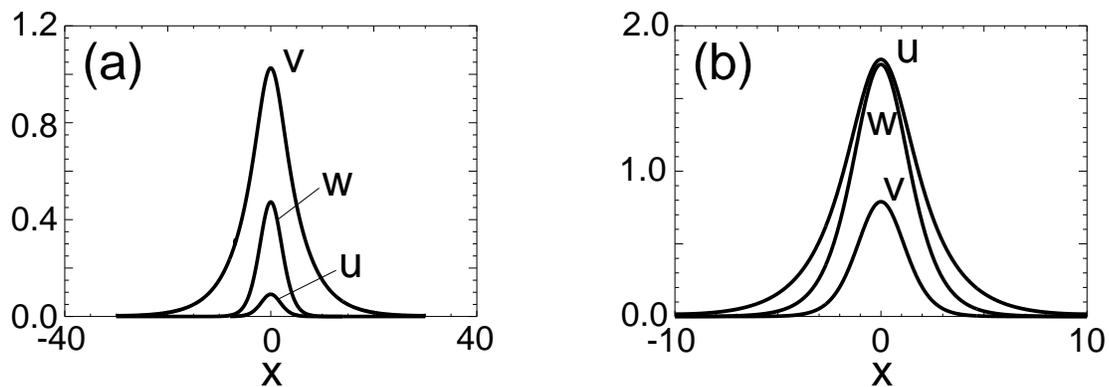}}
\caption{Examples of three-wave solitary waves of the model (\ref{normal})
for (a) $\alpha = 0.05$,  $\alpha_1 =5$, and (b) $\alpha =5$, $\alpha_1
=0.35$.}
\end{figure}

\begin{figure}
\setlength{\epsfxsize}{15.0cm}
\mbox{\epsffile{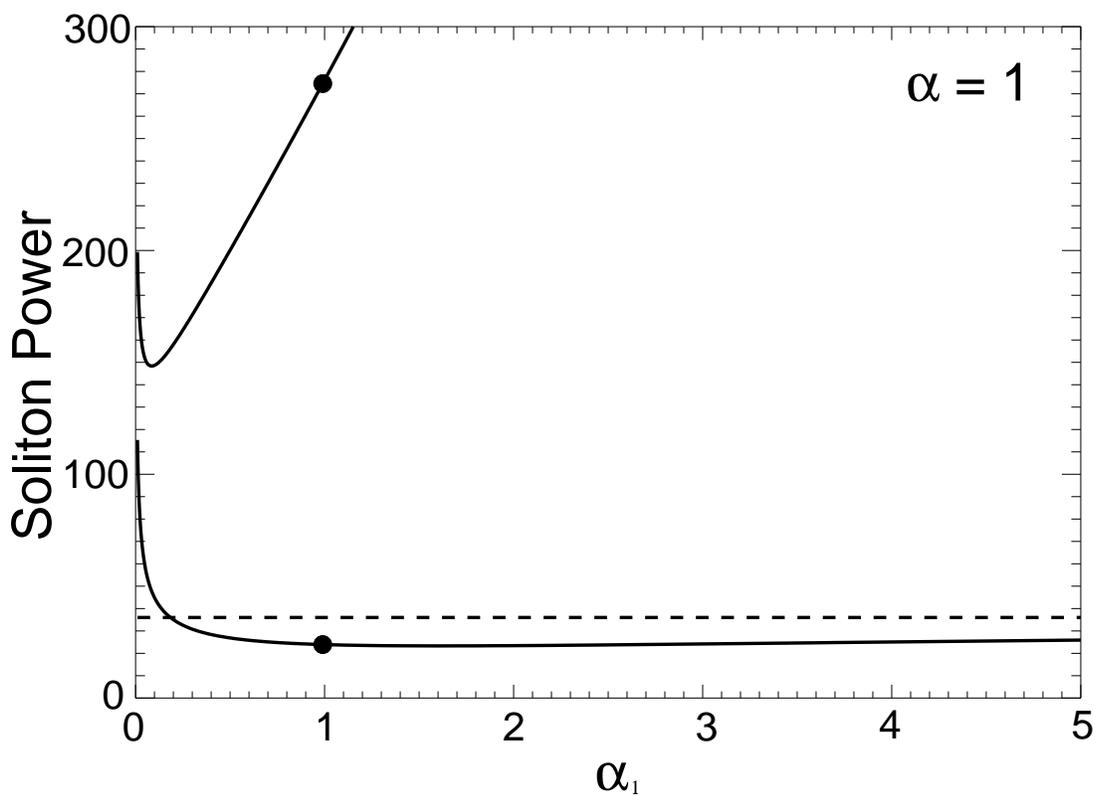}}
\caption{
Two branches of multistep cascading solitons shown as the total
soliton power $P$ vs. the parameter $\alpha_1$ for $\alpha =1$ and $\chi
=1$. Filled circles show the analytical solutions. The lower branch
approaches a family of two-wave quadratic solitons (for $\alpha_1
\rightarrow \infty$) shown by a dashed line.}
\end{figure}

\begin{figure}
\setlength{\epsfxsize}{15.0cm}
\mbox{\epsffile{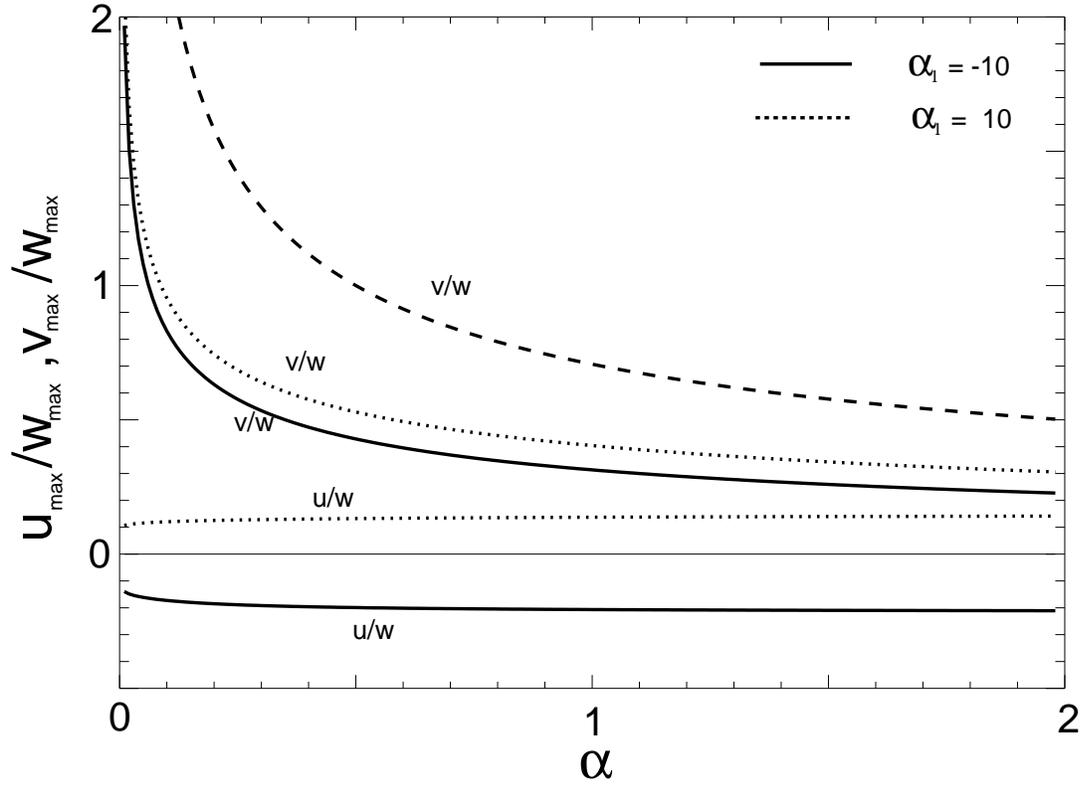}}
\caption{
Peak intensity ratios for the family of three-wave solitary waves
($\chi =1$) at  $\alpha_1 = -10$ (solid), $\alpha_1 = 10$ (doted). Upper
dashed curve shows the asymptotic limit of large $\alpha_1$ corresponding
to the two-wave quadratic solitons.}
\end{figure}

\begin{figure}
\setlength{\epsfxsize}{15.0cm}
\mbox{\epsffile{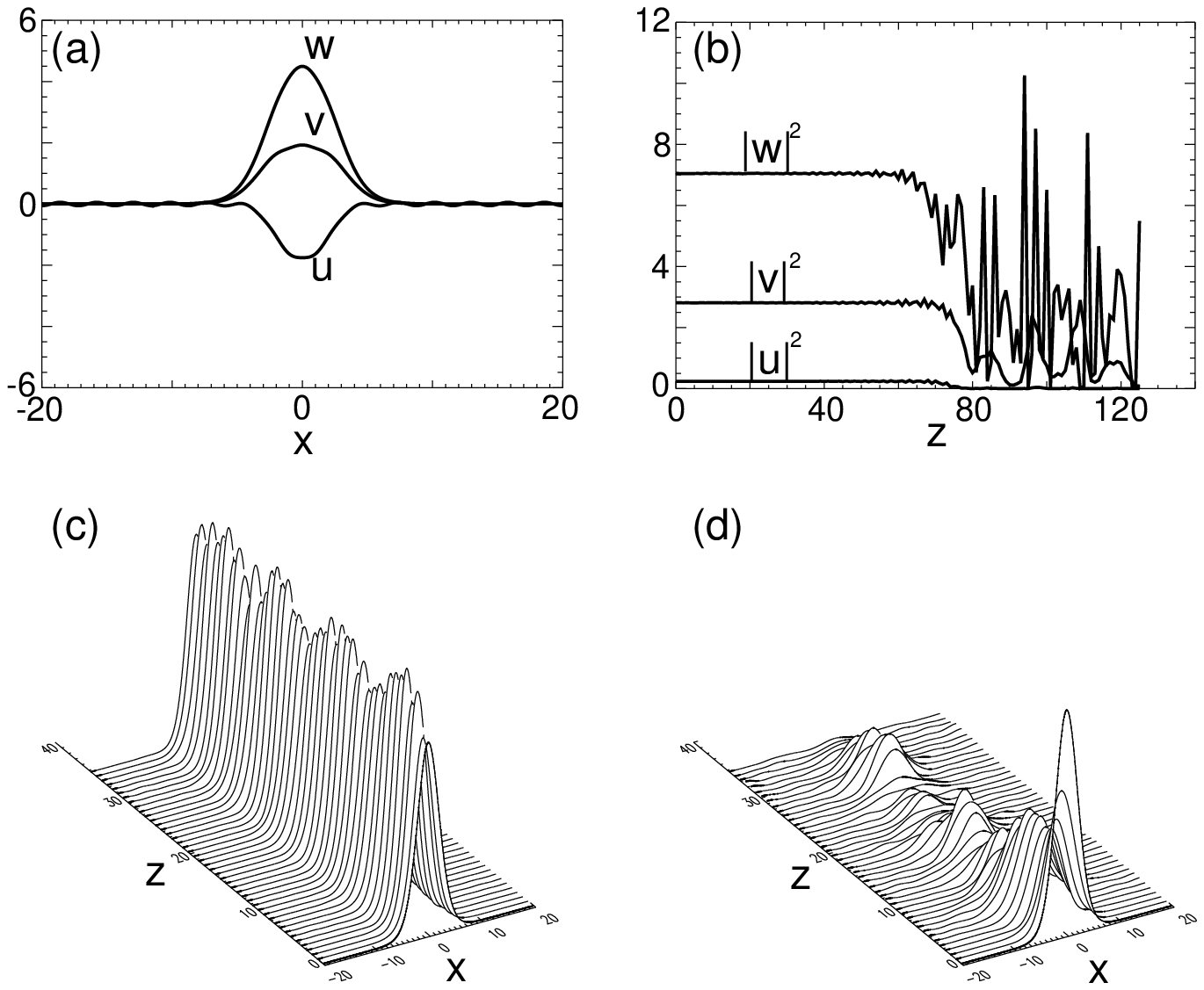}}
\caption{
Examples of (a) a three-wave quasi-soliton ($\alpha=1$,
$\alpha_1=-5$) and (b) its instability-induced long-term evolution shown
for  the peak intensities ($\alpha =1$, $\alpha_1 = -10$). (c,d) Evolution
of the fundamental harmonic of a two-wave soliton owing to an unseeded
unmatched SFM process with  $\alpha_1 = +8$ and $\alpha_1 =  -8$,
respectively.}
\end{figure}


\begin{references}

\bibitem{steg} For an overview, see G. Stegeman, D.J. Hagan, and L. Torner,
Opt. Quantum Electron. {\bf 28}, 1691 (1996);  Yu.S. Kivshar, In: {\em
Advanced Photonics with Second-Order Optically Nonlinear Processes}, Eds.
A.D. Boardman {\em et al.} (Kluwer,  Amsterdam, 1998), p. 451.

\bibitem{pelin} D.E. Pelinovsky, A.V. Buryak, and Yu.S. Kivshar, Phys. Rev.
Lett. {\bf 75}, 591 (1995); L. Torner, D. Michalache, D. Mazilu, and N.
Akhmediev, Opt. Lett. {\bf 20}, 2183 (1995).

\bibitem{koy_sal} K. Koynov and S. Saltiel, Opt. Commun. {\bf 152},
96 (1998).

\bibitem{akhmanov} See, e.g., S. Akhmanov, A. Dubovik, S. Saltiel,
I. Tomov, and V. Tunkin, JETP Lett. {\bf 20}, 117 (1974).

\bibitem{ol_bur} A.V. Buryak and Yu.S. Kivshar, Opt. Lett. {\bf 20}, 1612
(1994);  Phys. Lett. A {\bf 197}, 407 (1995).

\bibitem{pfister} O. Pfister, J.S. Wells, L. Hollberg, L. Zink, D.A. Van
Baak, M.D. Levenson, and W.R. Basenberg, Opt. Lett. {\bf 22}, 1211 (1997).

\bibitem{baldi} P. Baldi, C.G. Trevino-Palacios, G.I. Stegeman, M.P. De
Micheli, D.B. Ostrowsky, D. Delacourt, and M. Papuchon, Electron. Lett.
{\bf 31}, 1350 (1995).

\bibitem{malomed} B.A. Malomed, D. Anderson, and M. Lisak, Opt. Commun.
{\bf 126}, 251 (1996).

\bibitem{trillo} A.V. Buryak, Yu.S. Kivshar, and S. Trillo,  Phys. Rev.
Lett. {\bf 77}, 5210 (1996).

\bibitem{zakh} V.E. Zakharov and E.A. Kuznetsov, Zh. Eksp. Teor. Fiz.
{\bf 113}, 1892 (1998) [JETP {\bf 86}, 1035 (1998)].

\bibitem{KdV} See, e.g., V.I. Karpman, Phys. Lett. A {\bf 193}, 355 (1994).

\end{references}
\end{document}